\documentclass[aps,preprint,showpacs,showkeys]{revtex4}
\usepackage{amsfonts}
\usepackage{amsmath}
\usepackage{amssymb}
\usepackage{graphicx}

\begin{document}

\preprint{HEP/123-qed}
\title[Short title for running header]{Geometrical properties of Riemannian
superspaces, observables and physical states.\footnote{To Anna Grigorievna Kartavenko and Academic Professor Alexei Norianovich Sissakian, {\em in memoriam}}}
\author{Diego Julio Cirilo-Lombardo}
\affiliation{International Institute of Physics, Natal, RN, Brazil}
\affiliation{Bogoliubov Laboratory of Theoretical Physics Joint Institute for Nuclear
Research, 141980, Dubna, Russian Federation}
\pacs{PACS number}

\begin{abstract}
Classical and quantum aspects of physical systems that can be described by
Riemannian non degenerate superspaces are analyzed from the topological and
geometrical points of view. For the $N=1$ case the simplest supermetric
introduced in [Physics Letters B \textbf{661}, (2008),186] have the correct
number of degrees of freedom for the fermion fields and the super-momentum
fulfil the mass shell condition, in sharp contrast with other cases in the
literature where the supermetric is degenerate. This fact leads a deviation
of the 4-impulse (e.g. mass constraint) that can be mechanically interpreted
as a modification of the Newton's law. Quantum aspects of the physical
states and the basic states and the projection relation between them, are
completely described due the introduction of a new Majorana-Weyl
representation of the generators of the underlying group manifold. A new
oscillatory fermionic effect in the $B_{0}$ part of the vaccum solution
involving the chiral and antichiral components of this Majorana bispinor is
explicitly shown.
\end{abstract}

\email{diego@iip.ufrn.br\\
diego@theor.jinr.ru}
\maketitle
\tableofcontents

\section{Introduction}

The problem of giving an unambiguous quantum mechanical description of a
particle in a general spacetime has been repeatedly investigated. The
introduction of supersymmetry provided a new approach to this question,
however, some important aspects concerning the physical observables remain
not completely understood, classically and quantically speaking.

The superspace concept, on the other hand, simplify considerably the link
between ordinary relativistic systems and ``supersystems", extending an
standard (bosonic) spacetime by means of a general (super)group manifold,
equipped with also fermionic (odd) coordinates.

In a previous reference [2] we introduced, besides other supersymmetric
quantum systems of physical interest, a particular $N=1$ superspace [3],
with the aim of studying a superworld-line quantum particle (analogously to
the relativistic one) and its relation with SUGRA theories [1,2,10]. The
main feature of this superspace is that the supermetric is \emph{invertible
and non-degenerate}, and is the basic ingredient of a Volkov-Pashnev
particle action [3 ] that is, of type $G4$ in the Casalbuoni's
classification [4]. As shown in [2,10], the non-degeneracy of the
supermetrics (and therefore the corresponding superspaces) in the
description of physical systems leads to important geometrical and
topological effects on the quantum states, namely, \textit{consistent
mechanisms of localization and confinement}, due purely to the geometrical
character of the Lagrangian. Also an \textit{alternative to the
Randall-Sundrum(RS)} model without extra bosonic coordinates can be
consistently formulated, eliminating the problems that the RS-like models
present at the quantum level[2].

Given the importance of the non degeneracy of the supermetrics in the
formulation of physical theories, in the present letter we analyze further
the super-line element introduced in [2,3] in several senses. In Section II,
classical aspects of physical observables arising from the geometrical
properties of the superspace are derived, analyzed and compared with the
results of well known models where the metric is degenerate. Section III\ is
devoted to describe the bosonic $B_{0}$ part of the solution states and the
corresponding probability currents ($j_{0}$), in the temporal and
geometrical degeneracy limits ($t,a\rightarrow 0,\infty $). In Section IV, a 
\textit{new Majorana-Weyl representation} for the generators of the
metaplectic group[9] is introduced. The reduction from the 6 dimensional
group manifold to the 4 dimensional spacetime is explicitly given and the
specific action of the generators of the metaplectic group on the physical
and basic states, is performed. Finally, in Section V, the concluding
remarks and comments on the main results are given.

\section{Fermionic momentum and supervelocity: physical constraints on the
superparticle action}

To begin with, we consider the line element of the non-degenerated
supermetric [2,3] introduced in \cite{diego1} 
\begin{equation}
ds^{2}=\omega ^{\mu }\omega _{\mu }+{\mathbf{a}}\omega ^{\alpha }\omega
_{\alpha }-{\mathbf{a}}^{\ast }\omega ^{\dot{\alpha}}\omega _{\dot{\alpha}},
\end{equation}%
where the bosonic term and the Majorana bispinor compose a superspace $%
(1,3|1)$, with coordinates $(t,x^{i},\theta ^{\alpha },\bar{\theta}^{\dot{%
\alpha}})$, and where Cartan forms of supersymmetry group are described by 
\begin{equation}
\omega _{\mu }=dx_{\mu }-i(d\theta \sigma _{\mu }\bar{\theta}-\theta \sigma
_{\mu }d\bar{\theta}),\qquad \omega ^{\alpha }=d\theta ^{\alpha },\qquad
\omega ^{\dot{\alpha}}=d\theta ^{\dot{\alpha}}
\end{equation}%
(obeying evident supertranslational invariance). As we have extended our
manifold to include fermionic coordinates, it is natural to extend also the
concept of trajectory of point particle to the superspace. To do this we
take the coordinates $x\left( \tau \right) $, $\theta ^{\alpha }\left( \tau
\right) $ and $\overline{\theta }^{\overset{.}{\alpha }}\left( \tau \right) $
depending on the evolution parameter $\tau .$ Geometrically, the action
functional that will describe the world-line of the superparticle reads 
\begin{equation}
S=\int_{\tau _{1}}^{\tau _{2}}d\tau L\left( x,\theta ,\overline{\theta }%
\right) =-m\int_{\tau _{1}}^{\tau _{2}}d\tau \sqrt{\overset{\circ }{\omega
_{\mu }}\overset{\circ }{\omega ^{\mu }}+{\mathbf{a}}\overset{.}{\theta }%
^{\alpha }\overset{.}{\theta }_{\alpha }-{\mathbf{a}}^{\ast }\overset{.}{%
\overline{\theta }}^{\overset{.}{\alpha }}\overset{.}{\overline{\theta }}_{%
\overset{.}{\alpha }}}
\end{equation}%
where $\overset{\circ }{\omega _{\mu }}=\overset{.}{x}_{\mu }-i(\overset{.}{%
\theta }\ \sigma _{\mu }\overline{\theta }-\theta \ \sigma _{\mu }\overset{.}%
{\overline{\theta }})$, and the dot indicates derivative with respect to the
parameter $\tau $, as usual.

The momenta, canonically conjugated to the coordinates of the superparticle,
are 
\begin{equation}
\begin{array}{rcl}
\mathcal{P}_{\mu } & = & \partial L/\partial x^{\mu }=\left( m^{2}/L\right) 
\overset{\circ }{\omega _{\mu }} \\ 
\mathcal{P}_{\alpha } & = & \partial L/\overset{.}{\partial \theta ^{\alpha }%
}=i\mathcal{P}_{\mu }\left( \sigma ^{\mu }\right) _{\alpha \overset{.}{\beta 
}}\overline{\theta }^{\overset{.}{\beta }}+\left( m^{2}{\mathbf{a}}/L\right) 
\overset{.}{\theta _{\alpha }} \\ 
\mathcal{P}_{\overset{.}{\alpha }} & = & \partial L/\overset{.}{\partial 
\overline{\theta }^{\overset{.}{\alpha }}}=i\mathcal{P}_{\mu }\theta
^{\alpha }\left( \sigma ^{\mu }\right) _{\alpha \overset{.}{\alpha }}-\left(
m^{2}{\mathbf{a}}^{\ast }/L\right) \overset{.}{\overline{\theta }_{\overset{.%
}{\alpha }}}%
\end{array}%
\end{equation}

Notice the first important fact: the fermionic momenta $\mathcal{P}_{\alpha
} $ and $\mathcal{P}_{\overset{.}{\alpha }}$ are proportional to $\overset{.}%
{\theta _{\alpha }}$ and $\overset{.}{\overline{\theta }_{\overset{.}{\alpha 
}}}$ due to ${\mathbf{a}}$ and ${\mathbf{a}}^{\ast }$. That is not the case
in the standard superparticle actions where the superspace metric is
degenerate, \emph{e.g.} without these complex coefficients. This is the
case, for instance, of the constrained Brink-Schwartz superparticle model or
the relativistic $G4$ of Casalbuoni (see [4,5,6]), where the Lagrangian
presents only terms of the form $\sim \overline{\theta }^{\overset{.}{\alpha 
}}\overset{.}{\overline{\theta }}_{\overset{.}{\alpha }},\theta ^{\alpha }%
\overset{.}{\theta }_{\alpha }$ etc., which leads to fermionic momenta
proportional to the coordinates \emph{e.g.} $\mathcal{P}_{\overset{.}{\alpha 
}}=\partial L/\overset{.}{\partial \overline{\theta }^{\overset{.}{\alpha }}}%
\varpropto \overline{\theta }_{\overset{.}{\alpha }}$. This reduces,
consequently, the number of fermionic degrees of freedom to $n$ instead of
the $2n$ logically required.

It is difficult to study this system in a Hamiltonian formalism framework
because of the constraints and the nullification of the Hamiltonian. As the
action (3) is invariant under reparametrizations of the evolution parameter (%
$\tau \rightarrow \widetilde{\tau }=f\left( \tau \right) $), one way to
overcome this difficulty is to identify the dynamical variable $x_{0}$ with
the time. It is sufficient to introduce the concepts of integration and
derivation in supermanifolds, as we have done in \cite{diego1}, to have the
action rewritten in the form 
\begin{equation}
S=-m\int_{\tau 1}^{\tau 2}\overset{.}{x}_{0}d\tau \sqrt{\left[ 1-iW_{,0}^{0}%
\right] ^{2}-\left[ x_{,0}^{i}-W_{,0}^{i}\right] ^{2}+\overset{.}{x}%
_{0}^{-2}\left( {\mathbf{a}}\overset{.}{\theta _{\alpha }}\overset{.}{\theta
^{\alpha }}-{\mathbf{a}}^{\ast }\overset{.}{\overline{\theta }_{\overset{.}{%
\alpha }}}\overset{.}{\overline{\theta }^{\overset{.}{\alpha }}}\right) }
\end{equation}%
where the $W_{,0}^{\mu }$ are defined by 
\begin{equation}
\overset{\circ }{\omega }^{0}=\overset{.}{x}^{0}\left[ 1-iW_{,0}^{0}\right]
,\qquad \overset{\circ }{\omega }^{i}=\overset{.}{x}^{0}\left[
x_{,0}^{i}-iW_{,0}^{i}\right]
\end{equation}%
If $x_{0}\left( \tau \right) $ is taken to be the evolution parameter, then 
\begin{equation}
S=-m\int_{x_{0}\left( \tau _{1}\right) }^{x_{0}\left( \tau _{2}\right)
}dx_{0}\sqrt{\left[ 1-iW_{,0}^{0}\right] ^{2}-\left[ x_{,0}^{i}-W_{,0}^{i}%
\right] ^{2}+{\mathbf{a}}\overset{.}{\theta }^{\alpha }\overset{.}{\theta }%
_{\alpha }-{\mathbf{a}}^{\ast }\overset{.}{\overline{\theta }}^{\overset{.}{%
\alpha }}\overset{.}{\overline{\theta }}_{\overset{.}{\alpha }}}\equiv \int
dx_{0}L
\end{equation}%
Physically, this `dynamical parameter' $x_{0}$ corresponds to the time
measured by an observer's clock in the rest frame.

The total relativistic velocity in the superspace (supervelocity) can be
derived as usual from the line element of the supermetric, using a parameter
of evolution of the physical system $\tau $. Then we have, from (1), the
`true' supervelocity 
\begin{eqnarray}
\overset{\cdot }{z}^{A}\overset{\cdot }{z}_{A}\equiv v^{2}=\left( \frac{ds}{%
d\tau }\right) ^{2} &=&\dot{x}^{\mu }\dot{x}_{\mu }-2i\dot{x}^{\mu }(\dot{%
\theta}\sigma _{\mu }\bar{\theta}-\theta \sigma _{\mu }\dot{\bar{\theta}}) \\
&&+\left( {\mathbf{a-}}\bar{\theta}^{\dot{\alpha}}\bar{\theta}_{\dot{\alpha}%
}\right) \dot{\theta}^{\alpha }\dot{\theta}_{\alpha }-\left( {\mathbf{a}}%
^{\ast }+\theta ^{\alpha }\theta _{\alpha }\right) \dot{\bar{\theta}}^{\dot{%
\alpha}}\dot{\bar{\theta}}_{\dot{\alpha}}  \notag
\end{eqnarray}%
with $z_{A}\equiv \left( x_{\mu },\theta _{\alpha },\overline{\theta }_{%
\overset{\cdot }{\alpha }}\right) $.

Note that in the case of [4], due to the degeneracy of the superspace
metric, there is not supervelocity but the usual 4-velocity, as in the pure
relativistic case. There is also another notorious difference between the
4-velocity and the supervelocity, precisely coming from the classical
relativistic theory: in the relativistic case the 4-velocity fulfils $\left(
\hbar =c=1\right) $%
\begin{equation}
v_{rel}^{2}=\dot{x}^{\mu }\dot{x}_{\mu }=-1
\end{equation}%
in conicidence with the mass shell condition from the impulses $p_{\mu
}p^{\mu }=m^{2}$, as in the original $G4$ model [4].

In our case, due to the non degenerate super-line element (1), we have the
mass shell condition $p_{A}p^{A}=m^{2}$, with the $A$ index taking values on 
$(x,\theta ,\overline{\theta })$. Then, the `supervelocity' expression (8)
automatically fulfils%
\begin{eqnarray}
\overset{\cdot }{z}^{A}\overset{\cdot }{z}_{A}=-1 &=&\dot{x}^{\mu }\dot{x}%
_{\mu }-2i\dot{x}^{\mu }(\dot{\theta}\sigma _{\mu }\bar{\theta}-\theta
\sigma _{\mu }\dot{\bar{\theta}}) \\
&&+\left( {\mathbf{a-}}\bar{\theta}^{\dot{\alpha}}\bar{\theta}_{\dot{\alpha}%
}\right) \dot{\theta}^{\alpha }\dot{\theta}_{\alpha }-\left( {\mathbf{a}}%
^{\ast }+\theta ^{\alpha }\theta _{\alpha }\right) \dot{\bar{\theta}}^{\dot{%
\alpha}}\dot{\bar{\theta}}_{\dot{\alpha}}  \notag
\end{eqnarray}%
\emph{i.e.} our supersymmetric model is enforced to accomplish the standard
classical relativistic conditions.

\subsection{Some particular cases}

\begin{itemize}
\item[i)] It is not difficult to see that, in the case when the Majorana
spinors are null, or constant with respect to the evolution parameter of the
system (or proper time), we have $(c=1)$ 
\begin{equation}
\left. v^{2}\right\vert _{\theta ,\overline{\theta }=cte}=\dot{x}^{\mu }\dot{%
x}_{\mu }=-1.
\end{equation}%
That is, the system velocity is given by a pure bosonic contribution and we
recover the standard expression, due to (10). Notice that the relativistic
condition on the supervelocity will bring us several constraints between
bosonic and fermionic coordinates of the supersymmetric system.

\item[ii)] Now, if we consider the bosonic variables constant with respect
to the evolution parameter of the system, we arrive at 
\begin{equation}
\left. v^{2}\right\vert _{x=cte}=-1=\left( {\mathbf{a-}}\bar{\theta}^{\dot{%
\alpha}}\bar{\theta}_{\dot{\alpha}}\right) \dot{\theta}^{\alpha}\dot{\theta}%
_{\alpha}-\left( {\mathbf{a}}^{\ast}+\theta^{\alpha}\theta_{\alpha}\right) 
\dot{\bar{\theta}}^{\dot{\alpha}}\dot{\bar{\theta}}_{\dot{\alpha}},
\end{equation}
such that the kinetic term is due to Majorana bispinors. However, if the
time parameter is identified with $d\tau=dt$ (proper time), the relativistic
velocity for the static case (free fall) lead us to the following constraint 
\begin{eqnarray}
\left. v^{2}\right\vert _{\tau=x_{0}} & \rightarrow & 2i(\dot{\theta}%
\sigma_{0}\bar{\theta}-\theta\sigma_{0}\dot{\bar{\theta}})  \notag \\
& =&\left( {\mathbf{a-}}\bar{\theta}^{\dot{\alpha}}\bar{\theta}_{\dot{\alpha 
}}\right) \dot{\theta}^{\alpha}\dot{\theta}_{\alpha}-\left( {\mathbf{a}}%
^{\ast}+\theta^{\alpha}\theta_{\alpha}\right) \dot{\bar{\theta}}^{\dot{\alpha%
}}\dot{\bar{\theta}}_{\dot{\alpha}}.
\end{eqnarray}
\end{itemize}

In the case of Casalbuoni's superparticle action ($\mathbf{a}=\mathbf{a}%
^{\ast }=0$), the expression (10) is reduced to%
\begin{equation}
\left. v^{2}\right\vert _{\mathbf{a}=\mathbf{a}^{\ast }=0}=\left( \frac{ds}{%
d\tau }\right) ^{2}=\dot{x}^{\mu }\dot{x}_{\mu }-2i\dot{x}^{\mu }(\dot{\theta%
}\sigma _{\mu }\bar{\theta}-\theta \sigma _{\mu }\dot{\bar{\theta}})+\left( 
\bar{\theta}^{\dot{\alpha}}\bar{\theta}_{\dot{\alpha}}\right) (\dot{\theta}%
^{\alpha }\dot{\theta}_{\alpha })-\left( \theta ^{\alpha }\theta _{\alpha
}\right) (\dot{\bar{\theta}}^{\dot{\alpha}}\dot{\bar{\theta}}_{\dot{\alpha}})
\end{equation}%
Passing to the proper time and free fall cases as before, we arrive to the
important conclusion that in the Casalbuoni's model 
\begin{equation}
\left. v^{2}\right\vert _{\tau =x_{0}}\rightarrow 2i(\dot{\theta}\sigma _{0}%
\bar{\theta}-\theta \sigma _{0}\dot{\bar{\theta}})\equiv \left( \bar{\theta}%
^{\dot{\alpha}}\bar{\theta}_{\dot{\alpha}}\right) (\dot{\theta}^{\alpha }%
\dot{\theta}_{\alpha })-\left( \theta ^{\alpha }\theta _{\alpha }\right) (%
\dot{\bar{\theta}}^{\dot{\alpha}}\dot{\bar{\theta}}_{\dot{\alpha}})
\end{equation}%
In order to fulfil this condition, the fermions shall be null or constant
with respect to the evolution parameter of the system. This is a direct
consequence of the degeneration (non invertibility) of the supermetric in
the Brink-Schwartz or the Casalbuoni-$G4$ type cases. This inconsistence of
supersymmetric models based on degenerate supermetrics is translated into
the incompatibility of the `natural constraints' for the relativistic
conditions on the corresponding superline elements.

In the full super-relativistic case, however, we obtain the mass shell
condition $p_{A}p^{A}=m^{2}$, with $A=(x,\theta ,\overline{\theta })$. But,
in sharp contrast with the super-line element used by Casalbuoni, the square
of the 4-impulse is not $m^{2}$, as is easily seen from the expressions (4).
This issue will be treated deeply in [7].

In summary: non degenerate (invertible) supermetrics basis of relativistic $%
G4$ models lead to consistent SUSY analogs of (pseudo)classical relativistic
systems, in the sense that the phase space number of degrees of freedom is
the correct ones. Therefore, the exact quantization of the system can be
performed by standard procedures: canonical (Gupta-Bleuler in the case of
QFT, etc.)[3,6] or non-canonical ones (group theoretical, geometrical,
coherent states quantizations, etc.)[1,2,9].

Also the 4-momentum in the case of the non degenerate superspace does not
fulfil the mass shell condition. This fact can be (classically) translated
to an observable deviation from the Newton's law of gravitation. In the next
sections the meaning of these constraints and the possible explanation for
deviations of the GR predictions will be briefly discussed.

\section{On the $B_{0}$ part of the spinorial solution}

From previous works \cite{diego1, diego2}, the supermultiplet solution for
the geometric lagrangian is 
\begin{eqnarray}
g_{ab}(0,\lambda)&=&\left\langle \psi_{\lambda}\left( t\right) \right\vert
L_{ab}\left\vert \psi_{\lambda}\left( t\right) \right\rangle \\
&=&e^{-\left( \frac{m}{\left\vert a\right\vert }\right)
^{2}t^{2}+c_{1}t+c_{2}}e^{\xi\varrho\left( t\right)
}\chi_{f}\langle\psi_{\lambda}(0)|\left( 
\begin{array}{c}
a \\ 
a^{\dagger}%
\end{array}
\right) _{ab}|\psi_{\lambda}(0)\rangle  \notag
\end{eqnarray}
Consider, for simplicity, the `square' solution for the three compactified
dimensions \cite{diego2} (spin $\lambda$ fixed) 
\begin{equation}
g_{ab}(t)=e^{A(t)+\xi\left( \phi_{\alpha}(t)+\bar{\chi}_{\dot{\alpha}%
}(t)\right) }g_{ab}(0)
\end{equation}
We have obtained schematically for the exponential fermionic part 
\begin{eqnarray}
\varrho\left( t\right) \equiv\overset{\circ}{\phi}_{\alpha}\left[%
\left(\alpha e^{i\omega t/2}\right.\right. &\left.\left.+\right.\right.
&\left.\left. \beta e^{-i\omega t/2}\right)-\left(\sigma^{0}\right)_{\overset%
{.}{\alpha}}^{\alpha} \left(\alpha e^{i\omega t/2}-\beta e^{-i\omega
t/2}\right) \right]  \notag \\
&+& \frac{2i}{\omega}\left[ \left( \sigma^{0}\right) _{\alpha}^{\overset{.}{%
\ \beta}}\ \overline{Z}_{\overset{.}{\beta}}+\left( \sigma^{0}\right) _{\ 
\overset{.}{\alpha}}^{\alpha}\ Z_{\alpha}\right]
\end{eqnarray}

For the exponential bosonic part we have 
\begin{equation}
A(t)=-\left( \frac{m}{|{\mathbf{a}}|}\right) ^{2}t^{2}+c_{1}t+c_{2}
\end{equation}%
And the initial value for the metric is given by 
\begin{equation}
g_{ab}(0)=\langle \psi (0)|\left( 
\begin{array}{c}
a \\ 
a^{\dagger }%
\end{array}%
\right) _{ab}|\psi (0)\rangle ,
\end{equation}%
where $\overset{\circ }{\phi }_{\alpha },Z_{\alpha },\overline{Z}_{\overset{.%
}{\beta }}$ are constant spinors, and $\alpha $ and $\beta $ are $\mathbb{C}$%
-numbers. The constant $c_{1}\in \mathbb{C}$ due the obvious physical
reasons and the chirality restoration of the superfield solution [1,2,10].

Notice that there exists a factor in the vacuum solution eq.(16), that we
call $\chi _{f}$, coming from the odd generators of the big covering group
related to the symmetries of the specific model (3). These `parafermionic'
part and the associated odd generators [1,9,10] will not be treated in this
paper, and will be left aside.

Two geometric-physical options will be related to the orientability of the
superspace trajectory[11]: $\alpha =\pm \beta $. We take, without lose
generality $\alpha =+\beta $ then, exactly,%
\begin{equation}
\varrho \left( t\right) =\left( 
\begin{array}{c}
\overset{\circ }{\phi }_{\alpha }\cos \left( \omega t/2\right) +\frac{2}{%
\omega }Z_{\alpha } \\ 
-\overset{\circ }{\overline{\phi }}_{\overset{\cdot }{\alpha }}\sin \left(
\omega t/2\right) -\frac{2}{\omega }\overline{Z}_{\overset{.}{\alpha }}%
\end{array}%
\right)
\end{equation}%
which obviously represents a \textit{Majorana fermion} where the $\mathbb{C}$
symmetry is inside of the constant spinors.

The spinorial part of the superfield solution in the exponent becomes 
\begin{equation}
\xi\varrho\left( t\right) =\theta^{\alpha}\left( \overset{\circ}{\phi }%
_{\alpha}\cos\left( \omega t/2\right) +\frac{2}{\omega}Z_{\alpha}\right) -%
\overline{\theta}^{\overset{\cdot}{\alpha}}\left( -\overset{\circ}{\overline{%
\phi}}_{\overset{\cdot}{\alpha}}\sin\left( \omega t/2\right) -\frac{2}{\omega%
}\overline{Z}_{\overset{.}{\alpha}}\right)
\end{equation}

In the above expression there appear a type of Zitterbewegung or continous
oscillation between the chiral and antichiral part of the bispinor $\varrho
(t)$. The physical meaning of such an oscillation is not completely clear
for us given that is an effect never described before from the theoretical
point of view. Figures 1, 2 and 3 are describing qualitatively such effect
for suitable values of the parameters of the vacuum solution and with an
increasing $\omega $ respectively ($\omega _{1}<\omega _{2}<\omega _{3}$)
This important issue will be aborded in detail in a separate publication [8].

The interesting point of the physical states (and the basic ones) in
explicit form is that their behaviour (and the behavior of the zero
component of the current of probability $j_{0}(t)$), can be analyzed in the
limits of interest. This fact is very important in order to understand
deeply the even part $B_{0}$ of the superfield solution, then, the fermionic
and bosonic evolution of the system.

\subsection{Temporal limits}

i) For the square states (observables) 
\begin{equation}
g_{ab}(t)=\overset{F\left( t\right) }{\overbrace{e^{-\left( \left( \frac{m}{%
|a|}\right) ^{2}t^{2}+c_{1}t+c_{2}\right) }e^{\xi \rho (t)}}}\overset{v_{ab}}%
{\overbrace{\left[ \alpha \left( 
\begin{array}{c}
1 \\ 
0%
\end{array}%
\right) +\alpha ^{\ast }\left( 
\begin{array}{c}
0 \\ 
1%
\end{array}%
\right) \right] }}
\end{equation}%
the standard temporal limits at $t=0$ and $t\rightarrow \infty $ are 
\begin{equation}
g_{ab}(t=0)=F(0)v_{ab},\qquad F(0)=e^{-c_{2}}e^{\theta ^{\alpha }\overset{%
\circ }{\phi }_{\alpha }+\frac{2}{\omega }\left( \theta ^{\alpha }Z_{\alpha
}+\overline{\theta }^{\overset{\cdot }{\alpha }}\overline{Z}_{\overset{.}{%
\alpha }}\right) }
\end{equation}%
and 
\begin{equation}
g_{ab}(t\rightarrow \infty )\rightarrow 0
\end{equation}

The zero component of the current of probability for this observable reads 
\begin{equation}
j_{0}(t)=2E^{2}g_{ab}^{\dagger }(t)g^{ab}(t)
\end{equation}%
and therefore is positive definite (the energy appears squared), as was
pointed out in \cite{diego1, diego2}. The corresponding temporal limits at $%
t=0$ and $t\rightarrow \infty $ are 
\begin{equation}
j(t=0)=2E^{2}g_{ab}^{\dagger }(0)g^{ab}(0)=4E^{2}|\alpha |^{2}F^{2}(0)
\end{equation}%
and 
\begin{equation}
j_{0}(t\rightarrow \infty )\rightarrow 0
\end{equation}%
ii) Analogously, for the `square root' basic (non-observable) states we have 
\begin{equation}
\psi (t)=\overset{\equiv F^{1/2}\left( t\right) }{\overbrace{e^{-\frac{1}{2}%
\left( \left( \frac{m}{|a|}\right) ^{2}t^{2}+c_{1}t+c_{2}\right) }e^{\frac{1%
}{2}\xi \rho (t)}}}\overset{\equiv v_{ab}^{1/2}}{\overbrace{\left[ \sqrt{%
\alpha }\left( 
\begin{array}{c}
1 \\ 
0%
\end{array}%
\right) +\sqrt{\alpha ^{\ast }}\left( 
\begin{array}{c}
0 \\ 
1%
\end{array}%
\right) \right] }}
\end{equation}%
The standard temporal limit at $t=0$ takes the form 
\begin{equation}
\psi (0)=F^{1/2}(0)v_{ab}^{1/2}\qquad \text{with}\quad F^{1/2}(0)=e^{-\frac{%
c_{2}}{2}}e^{\frac{\theta ^{\alpha }\overset{\circ }{\phi }_{\alpha }}{2}+%
\frac{1}{\omega }\left( \theta ^{\alpha }Z_{\alpha }+\overline{\theta }^{%
\overset{\cdot }{\alpha }}\overline{Z}_{\overset{.}{\alpha }}\right) }
\end{equation}%
And, at $t\rightarrow \infty $, is 
\begin{equation}
\psi (t\rightarrow \infty )\rightarrow 0
\end{equation}

The current of probability for the `square root' states (non-observable) 
\begin{equation}
j_{0}(t)=2E\psi^{\dagger}(t)\psi(t)
\end{equation}
Then, as was pointed out in \cite{diego1, diego2}, it is \emph{not} positive
definite and the limits at $t=0$%
\begin{equation}
j_{0}(0)=2E\psi^{\dagger}(0)\psi(0)=4E|\alpha| F^{1/2}(0)
\end{equation}
and at $t\rightarrow\infty$%
\begin{equation}
j_{0}(t\rightarrow\infty) \rightarrow0
\end{equation}

\subsection{Special limits ($m$ fixed)}

\subsubsection{Physical states}

Two limiting cases are to be considered

\begin{itemize}
\item[i)] Bosonic ultralocalization: lim $a\rightarrow0$, $%
\omega\rightarrow\infty$, $\xi\varrho(t)\rightarrow0$%
\begin{equation}
g_{ab}(t)=\overset{F\left(t\right)}{\overbrace{e^{-\left( \frac{m}{|a|}%
\right) ^{2}t^{2}}}} v_{ab},\text{ \ \ \ \ \ \ } j_{0}(t)=4E^{2}|\alpha|^{2}
e^{-\frac{2m}{|a|}^{2}t^{2}}
\end{equation}
only boson part remains.

\item[ii)] Fermionic bosonization: lim $a\rightarrow \infty $, $\omega
\rightarrow 0$, $\xi \varrho \left( t\right) \rightarrow \theta ^{\alpha }%
\overset{\circ }{\phi }_{\alpha }+\frac{2}{\omega }\left( \theta ^{\alpha
}Z_{\alpha }+\overline{\theta }^{\overset{\cdot }{\alpha }}\overline{Z}_{%
\overset{.}{\alpha }}\right) $ 
\begin{equation}
g_{ab}(t)=\overset{F\left( t\right) }{\overbrace{e^{-\left(
c_{1}t+c_{2}\right) }e^{\theta ^{\alpha }\overset{\circ }{\phi }_{\alpha }+%
\frac{2}{\omega }\left( \theta ^{\alpha }Z_{\alpha }+\overline{\theta }^{%
\overset{\cdot }{\alpha }}\overline{Z}_{\overset{.}{\alpha }}\right) }}}%
v_{ab},\text{ \ \ \ \ \ \ }j_{0}(t)=4E^{2}|\alpha |^{2}e^{-2c_{2}}e^{2\theta
^{\alpha }\overset{\circ }{\phi }_{\alpha }+\frac{4}{\omega }\left( \theta
^{\alpha }Z_{\alpha }+\overline{\theta }^{\overset{\cdot }{\alpha }}%
\overline{Z}_{\overset{.}{\alpha }}\right) }
\end{equation}%
only bosonized fermionic part remains at this limit.
\end{itemize}

\subsubsection{Basic states}

\begin{itemize}
\item[i)] Bosonic ultralocalization: lim $a\rightarrow0$, $%
\omega\rightarrow\infty$, $\xi \varrho\left(t\right) \rightarrow0$ 
\begin{equation}
\psi(t)=e^{-\frac{1}{2}\left( \frac{m}{|a|}\right) ^{2}t^{2}} v_{ab}^{1/2},%
\text{ \ \ \ \ \ }j_{0}(t)=4E^{2}|\alpha|^{2}e^{-\frac {m}{|a|}^{2}t^{2}}
v_{ab}^{1/2}
\end{equation}

\item[ii)] Fermionic bosonization: Lim $a\rightarrow \infty $, $\omega
\rightarrow 0$, $\xi \varrho \left( t\right) \rightarrow 0$ 
\begin{eqnarray}
\psi (t) &=&e^{-\frac{\left( c_{1}t+c_{2}\right) }{2}}e^{\frac{\theta
^{\alpha }\overset{\circ }{\phi }_{\alpha }}{2}+\frac{1}{\omega }\left(
\theta ^{\alpha }Z_{\alpha }+\overline{\theta }^{\overset{\cdot }{\alpha }}%
\overline{Z}_{\overset{.}{\alpha }}\right) }v_{ab}^{1/2}, \\
j_{0}(t) &=&4E^{2}|\alpha |^{2}e^{-c_{2}}e^{\theta ^{\alpha }\overset{\circ }%
{\phi }_{\alpha }+\frac{2}{\omega }\left( \theta ^{\alpha }Z_{\alpha }+%
\overline{\theta }^{\overset{\cdot }{\alpha }}\overline{Z}_{\overset{.}{%
\alpha }}\right) }
\end{eqnarray}%
With similar comments than for the physical states. Also, the dynamics of
the system trivialize in the limit when $\frac{t}{\left\vert \mathbf{a}%
\right\vert }\symbol{126}\omega t\rightarrow $cons$\tan $t.
\end{itemize}

It is very important to note that we study specifically the corresponding
currents (for the square states and for the basic ones) because the number
operator will be related to the Hamiltonian operator for the basic states,
and for the square ones that in essence can be different [1,9,10] and main
task of a future work [7].

\section{Exotic Majorana representations and spinor transformations}

Because the underlying non compact symmetry of the Lagrangian corresponds to 
$Mp(4)$ (metaplectic group in four dimensions), we need to introduce a
suitable (matrix) representation of its irreducible part, namely $Mp(2)$:
the 3 dimensional metaplectic group. That is such because $Mp(4)$ splits
into 2 irreducible $Mp(2)$ subgroups. The $Mp(2)$ is the four covering of
the $L_{3}$ (the 3-dimensional Lorentz group or \textit{little group}) or
the double covering of $SL(2R)$.

The new Majorana-Weyl representation that we introduce below is given by the
2 by 2 following operators 
\begin{equation}
\sigma _{\alpha }=\left( 
\begin{array}{cc}
0 & 1 \\ 
1 & 0%
\end{array}%
\right) ,\quad \sigma _{\beta }=\left( 
\begin{array}{cc}
0 & -1 \\ 
1 & 0%
\end{array}%
\right) ,\quad \sigma _{\gamma }=\left( 
\begin{array}{cc}
1 & 0 \\ 
0 & -1%
\end{array}%
\right) ,
\end{equation}%
where the required condition over such matrices $\sigma _{\alpha }\wedge
\,\sigma _{\beta }=\sigma _{\gamma },$ $\sigma _{\beta }\wedge \,\sigma
_{\gamma }=\sigma _{\alpha }$ and $\sigma _{\gamma }\wedge \,\sigma _{\alpha
}=-\sigma _{\beta }$, evidently holds (Lie group) given the underlying
non-compact symmetry.

Notice, however, that the Majorana-Weyl representation (40) gives, in some
sense, the undotted description of the $Mp(4)$. The adjoint part (dotted) is
absolutely analog. Then, it is sufficient to analyze the undotted (dotted)
part.

Using the new representation given by the matrices (40), the underlying
symmetry of the bosonic (even) part of the supermultiplet system, can be
manifestly exposed as follows [1,10]: from the geometrical Lagrangian, we
can calculate the non-observable spinor field (analogouslly the respective
square $g_{ab}$)%
\begin{equation}
\begin{array}{c}
\psi (t)=\overset{\equiv F^{1/2}\left( t\right) }{\overbrace{e^{-\frac{1}{2}%
\left( \left( \frac{m}{|a|}\right) ^{2}t^{2}+c_{1}t+c_{2}\right) }e^{\frac{1%
}{2}\xi \rho (t)}}}\overset{\equiv v_{ab}^{1/2}}{\overbrace{\left[ \sqrt{%
\alpha }\left( 
\begin{array}{c}
1 \\ 
0%
\end{array}%
\right) +\sqrt{\alpha ^{\ast }}\left( 
\begin{array}{c}
0 \\ 
1%
\end{array}%
\right) \right] }}, \\ 
g_{ab}(t)=\overset{F\left( t\right) }{\overbrace{e^{-\left( \left( \frac{m}{%
|a|}\right) ^{2}t^{2}+c_{1}t+c_{2}\right) }e^{\xi \rho (t)}}}\overset{v_{ab}}%
{\overbrace{\left[ \alpha \left( 
\begin{array}{c}
1 \\ 
0%
\end{array}%
\right) +\alpha ^{\ast }\left( 
\begin{array}{c}
0 \\ 
1%
\end{array}%
\right) \right] }}%
\end{array}%
\end{equation}%
where dependence on time is in $F\left( t\right) $ (or $F^{1/2}\left(
t\right) $), and the $\mathbb{C}$ basic spinor was defined as $v_{ab}$ (or $%
v_{ab}^{1/2}$) (whose meaning in terms of the coherent states will be shown
soon).

Also, it is necessary to introduce here the second basic spinor%
\begin{equation}
\begin{array}{cc}
w_{ab}^{1/2}=\left[ \sqrt{\alpha }\left( 
\begin{array}{c}
1 \\ 
0%
\end{array}%
\right) -\sqrt{\alpha ^{\ast }}\left( 
\begin{array}{c}
0 \\ 
1%
\end{array}%
\right) \right] , & w_{ab}=\left[ \alpha \left( 
\begin{array}{c}
1 \\ 
0%
\end{array}%
\right) -\alpha ^{\ast }\left( 
\begin{array}{c}
0 \\ 
1%
\end{array}%
\right) \right]%
\end{array}%
\end{equation}%
which is needed for the full description below. Notice that, at this stage,
the only necessary spinors in order to describe the dynamical relation
between the group manifold, the spacetime and the related symmetries
involved, are $v_{ab}$ $\left( v_{ab}^{1/2}\right) $, $w_{ab}\left(
w_{ab}^{1/2}\right) $ and the generators (matrix representation) of the
metaplectic group (40).

The action of the Majorana Weyl matrices defined above will lead to%
\begin{eqnarray}
\sigma_{\alpha}\psi(t)&=&F^{1/2}\left( t\right) \left(
v_{ab}^{1/2}\right)^{\ast}, \\
\sigma_{\beta}\psi(t)&=&-F^{1/2}\left( t\right)
\left(w_{ab}^{1/2}\right)^{\ast}, \\
\sigma_{\gamma}\psi(t)&=&F^{1/2}\left(t\right) \left( w_{ab}^{1/2}\right)
\end{eqnarray}

The projection relations can be put in a more compact form%
\begin{eqnarray}
\psi ^{\ast }\sigma _{\alpha }\psi &=&F\left( t\right) 2Re(\alpha ), \\
\psi ^{\ast }\sigma _{\beta }\psi &=&F\left( t\right) 2 Im(\alpha ), \\
\psi ^{\ast }\sigma _{\gamma }\psi &=&0. \\
|\psi |^{2} &=&F\left( t\right) 2|\alpha |.
\end{eqnarray}

\subsection{Group manifold $\rightarrow$ space-time reduction}

Using the ingredients above, the projections are reduced to simple
expressions that bring us the specific splitting of the eigenvalue of the $a$
operator ($q$ and $p$) 
\begin{eqnarray}
\psi ^{\ast }\sigma _{\alpha }\psi &=&|\psi |^{2}\frac{Re(\alpha )}{|\alpha |%
}\rightarrow \text{position} \\
\psi ^{\ast }\sigma _{\beta }\psi &=&|\psi |^{2}\frac{Im(\alpha )}{|\alpha |}%
\rightarrow \text{momentum} \\
\psi ^{\ast }\sigma _{\gamma }\psi &=&0.
\end{eqnarray}%
By means of the basic coherent states, the generator of the 6-dimensional $%
Mp(4)$ breaks down to the 4-dimensional (observable) spacetime.%
\begin{equation*}
\begin{tabular}{|l|l|l|}
\cline{1-1}\cline{3-3}
$Mp(4)$ & $\quad \underset{\text{ non-observable }}{\overset{\psi }{%
\longrightarrow }}\quad $ & $\underset{\text{ observable(physical)}}{g_{ab}}$
\\ \cline{1-1}\cline{3-3}
\end{tabular}%
\ 
\end{equation*}%
And this is not trivial: as explained in several references [1,9,10 and
references therein] the physical states are the projections of the
generators of the 6-dimensional $Mp(4)$ on the 4-dimensional spacetime by
means of the (unnobservable) coherent states of fractional spin. And the
normalized spinors have a total projection in terms of normalized coherent
states 
\begin{equation}
\frac{\psi ^{\ast }\vec{\sigma}\psi }{|\psi |^{2}}=\frac{\alpha }{|\alpha |}.
\end{equation}

However, in the general case the polar decomposition can be introduced as $%
\alpha =|\alpha |e^{i\theta }$, picking up the explicit phase of the
coherent state (and having precisely the square of the norm of the state
such an order parameter) 
\begin{equation}
\frac{\psi ^{\ast }\vec{\sigma}\psi }{|\psi |^{2}}=e^{i\theta }.
\end{equation}

\subsection{Symmetry transformations}

We now use the faithful real representation of the matrices previously
introduced as generators of the symmetry transformations. From the Majorana
representation of the operators and the previous results we have in matrix
form 
\begin{eqnarray}
g_{ab}\rightarrow \left. g_{ab}\right\vert _{i} &=&e^{-i\sigma ^{i}\varphi
_{i}(t)}g_{ab} \\
\psi \rightarrow \left. \psi ^{\mu }\right\vert _{i} &=&e^{-i\sigma
^{i}\varphi _{i}(t)}\psi
\end{eqnarray}%
where $\varphi _{i}(t)$ is the generic parameter associated to the action of
the generators (40) given the specific rotation (or boost). They can be
labeled $\alpha \left( t\right) $, $\beta \left( t\right) $ and $\gamma
\left( t\right) $.with the corresponding symmetry of the associated
generator.

We can explicitly write expressions above in the following way 
\begin{eqnarray}
\left. g_{ab}\right\vert _{\alpha } &=&\cos \left( \alpha (t)\right)
g_{ab}-i\sin \left( \alpha (t)\right) \sigma ^{\mu }g_{ab} \\
\left. \psi \right\vert _{\alpha } &=&\cos \left( \alpha (t)\right) \psi
-i\sin \left( \alpha (t)\right) \sigma ^{\mu }\psi .
\end{eqnarray}

and similarly for $\gamma .$ On the other hand, as $\sigma ^{\beta }$ has an
anti hermitian representation, it acts as a boost (hyperbolic rotation) 
\begin{eqnarray}
\left. g_{ab}\right\vert _{\beta } &=&\cosh \left( \beta (t)\right)
g_{ab}-\sinh \left( \beta (t)\right) \sigma ^{\mu }g_{ab} \\
\left. \psi \right\vert _{\beta } &=&\cosh \left( \beta (t)\right) \psi
-\sinh \left( \beta (t)\right) \sigma ^{\mu }\psi
\end{eqnarray}%
Then, we can consider the observable state affected with an arbitrary phase 
\begin{equation}
g_{ab}(t)=e^{-i\varphi (t)}F\left( t\right) v_{ab}
\end{equation}%
the square (observable) state under the action of $\sigma ^{\alpha }$
oscillates harmonically with respect of the "natural complex spinors"$v_{ab}$
and $w_{ab}$ as 
\begin{equation}
g_{ab}^{(1)}(t)=F\left( t\right) \left[ \cos \left( \alpha (t)\right)
v_{ab}-i\sin \left( \alpha (t)\right) v_{ab}^{\ast }\right]
\end{equation}%
the square (observable) under the action of $\sigma ^{\beta }$ is "boosted"
(hyperbolic rotation) as 
\begin{equation}
g_{ab}^{(2)}(t)=F\left( t\right) \left[ \cosh \left( \beta (t)\right)
v_{ab}-\sinh \left( \beta (t)\right) w_{ab}^{\ast }\right]  \notag
\end{equation}

Finally under the action of $\sigma ^{\gamma }$ we have 
\begin{equation}
g_{ab}^{(3)}(t)=F\left( t\right) \left[ \cos \left( \gamma (t)\right)
v_{ab}-i\sin \left( \gamma (t)\right) w_{ab}\right]  \notag
\end{equation}

Similar action of the transformation operators can be easily realized over
the square root states: we only need to change $v_{ab}\rightarrow
v_{ab}^{1/2}$, etc. Here is quite evident the important fundamental role of
the"natural complex spinors"$v_{ab}$ and $w_{ab}:$ all the action of the
group generators of the underlying maniflod over the states are clearly
described only in this preferred spinor basis,(i.e. eqs.(41,42)).

\section{Concluding remarks}

In this paper we have analyzed several general aspects of the model
introduced by Volkov and Pashnev in [3], and correctly interpreted from the
quantum and field theoretical points of view in [1,2,10]. The model is
characterized by a $N=1$ superspace equipped with a non-degenerate
supermetric.

We explicitly shown that, in full agreement with the Casalbuoni's $G4$ model
[4], the Lagrangian must be of the form of a measure: \emph{i.e.}, the
square root of the super-line element, in order to preserv, physically and
mathematically speaking, the classical relativistic symmetries (as in the
case of the classical relativistic particle), and also the fact that the
supersymmetry appears as a pure relativistic effect[4].

From the geometrical point of view, the super-line element (non-degenerate
supermetric) induces the square of the supervelocity that must be
constrained to $-c^{2}$ (as in the classical relativistic case) considering
the square of the supermomentum that fulfil automatically the mass shell
condition. This fact leads to a deviation of the 4-impulse (mass constraint)
that can be mechanically interpreted as a modification of the Newton's law.
The possibility of antigravitational effects, alternatives to dark matter
and the relation with the MOND conjecture and the conservation of the CPT\
symmetry in such a case is under study [7].

With respect to the solutions obtained, we have confirmed from another point
of view the results of refs.[1,2,10] that there are two types of states: the
basic (non-observable) ones and observable physical states. The basic states
are coherent states corresponding to the double covering of the $SL(2C)$ or
the metaplectic group [1,2,9,10] responsible for projecting the symmetries
of the 6 dimensional $Mp(4)$ group space to the 4 dimensional spacetime by
means of a bilinear combination of the $Mp(4)$ generators.

An important new result we have found is that there exist an oscillatory
fermionic effect in the $B_{0}$ part of the supermultiplet as a
Zitterwebegung, but between the chiral and antichiral components of this
Majorana bispinor. This effect has been, as far as we know, never mentioned
in the literature. This issue will be treated in detail in a future work
[7,8].

The temporal and special limits were explicitly expressed through the
observable and non-observable states and their corresponding probability
currents. Again, as pointed out in ref.[2], the $\left\vert a\right\vert $
plays the main role in the restoration of the chiral symmetry, as is easily
seen from the boson ultralocalization limit and the fermionic bosonization
limits. . Quantum aspects of the physical states, the basic ones and the
projection relation between them, were completely described due the
introduction of a new Majorana-Weyl representation of the generators of the
underlying group manifold. Also we find that the action of these generators
over the states vaccum solution of the model, are clearly described only in
a preferred spinor basis given by eqs.(41,42).

\section{Acknowledgements}

Many thanks are given to Professors Yu. P. Stepanovsky, John Klauder and E.
C. G. Sudarshan for their interest; and particularly to Dr, Victor I. Afonso
for several discussions in the subject and help me in the preparation of
this text. This work is in memory of Anna Grigorievna Kartavenko, one of the
main responsables of the International Department of the Joint Institute of
Nuclear Research, that pass away recently. She was as part of my family in
Dubna: an angel that help me in several troubles 

The author is partially supported by CNPQ- brazilian funds.

[5] L. Brink and J. H. Schwartz, Phys. Lett. B, \textbf{100 }(1981), 310.

[6] W. Siegel, Phys. Lett. B, \textbf{203} (1988), 79.

[7] D. J. Cirilo-Lombardo and V. I. Afonso, work in preparation.

[8] D.J. Cirilo-Lombardo, work in progress.

[9] T. F. Jordan, N. Mukunda and S. V. Pepper, J. Math. Phys. \textbf{4},
(1963), 1089.

[10] D. J. Cirilo-Lombardo, Foundations of Physics, \textbf{39} (2009) 373.

[11] Victor D Gershun and Diego Julio Cirilo-Lombardo, J. Phys. A\textbf{\ 43%
} (2010), 305401.

\end{document}